\PassOptionsToPackage{unicode}{hyperref}
\PassOptionsToPackage{hyphens}{url}
\documentclass[11pt]{article}
\usepackage{amsmath,amssymb}
\usepackage{lmodern}
\usepackage{iftex}
\ifPDFTeX
  \usepackage[T1]{fontenc}
  \usepackage[utf8]{inputenc}
  \usepackage{textcomp} % provide euro and other symbols
\else % if luatex or xetex
  \usepackage{unicode-math}
  \defaultfontfeatures{Scale=MatchLowercase}
  \defaultfontfeatures[\rmfamily]{Ligatures=TeX,Scale=1}
\fi
\IfFileExists{upquote.sty}{\usepackage{upquote}}{}
\IfFileExists{microtype.sty}{% use microtype if available
  \usepackage[]{microtype}
  \UseMicrotypeSet[protrusion]{basicmath} % disable protrusion for tt fonts
}{}
\makeatletter
\@ifundefined{KOMAClassName}{% if non-KOMA class
  \IfFileExists{parskip.sty}{%
    \usepackage{parskip}
  }{% else
    \setlength{\parindent}{0pt}
    \setlength{\parskip}{6pt plus 2pt minus 1pt}}
}{% if KOMA class
  \KOMAoptions{parskip=half}}
\makeatother
\usepackage{xcolor}
\usepackage{longtable,booktabs,array}
\usepackage{calc} % for calculating minipage widths
\usepackage{etoolbox}
\makeatletter
\patchcmd\longtable{\par}{\if@noskipsec\mbox{}\fi\par}{}{}
\makeatother
\IfFileExists{footnotehyper.sty}{\usepackage{footnotehyper}}{\usepackage{footnote}}
\makesavenoteenv{longtable}
\usepackage{graphicx}
\makeatletter
\def\maxwidth{\ifdim\Gin@nat@width>\linewidth\linewidth\else\Gin@nat@width\fi}
\def\maxheight{\ifdim\Gin@nat@height>\textheight\textheight\else\Gin@nat@height\fi}
\makeatother
\setkeys{Gin}{width=\maxwidth,height=\maxheight,keepaspectratio}
\makeatletter
\def\fps@figure{htbp}
\makeatother
\ifLuaTeX
  \usepackage{selnolig}  % disable illegal ligatures
\fi
\IfFileExists{bookmark.sty}{\usepackage{bookmark}}{\usepackage{hyperref}}
\IfFileExists{xurl.sty}{\usepackage{xurl}}{} % add URL line breaks if available
\hypersetup{
  hidelinks,
  pdfcreator={LaTeX via pandoc}}

\title{The impacts of human-cobot collaboration on perceived cognitive
  load and usability during an industrial task: an exploratory experiment}

\author{Étienne Fournier\textsuperscript{a\footnote{Contact author: \url{etienne.fournier@univ-grenoble-alpes.fr}}}, Dorilys Kilgus\textsuperscript{b}, Aurélie Landry\textsuperscript{a c},
Belal Hmedan\textsuperscript{c}, Damien Pellier\textsuperscript{c}, \\
Humbert Fiorino\textsuperscript{c}, Christine Jeoffrion\textsuperscript{a}\\
\textsuperscript{a} Univ. Grenoble Alpes \-- LIP PC2S, BP 47 - 38040 Grenoble Cedex 9 - France\\
\textsuperscript{b} Univ. Grenoble Alpes \-- LIG, BP 47 - 38040 Grenoble Cedex 9 - France\\
}
\date{}

\begin{document}

\maketitle

\begin{abstract}
  Since cobots (collaborative robots) are increasingly being introduced in
  industrial environments, being aware of their potential positive and
  negative impacts on human collaborators is essential. This study guides
  occupational health workers by identifying the potential gains (reduced
  perceived time demand, number of gestures and number of errors) and
  concerns (the cobot takes a long time to perceive its environment, which
  leads to an increased completion time) associated with working with
  cobots. In our study, the collaboration between human and cobot during
  an assembly task did not negatively impact perceived cognitive load,
  increased completion time (but decreased perceived time demand), and
  decreased the number of gestures performed by participants and the
  number of errors made. Thus, performing the task in collaboration with a
  cobot improved the user's experience and performance, except for
  completion time, which increased. This study opens up avenues to
  investigate how to improve cobots to ensure the usability of the
  human-machine system at work.

  %Background: Industry 4.0 implements smart technologies to increase
  %productivity and to decrease the associated risks. Using cobots is
  %considered by industry as a potential means to reduce physical
  %constraints and improve performance without having to replace the human
  %factor. Research has yet to prove these benefits on humans.

  %Purpose: The goal of this study was to determine if working with a cobot
  %improved perceived cognitive load of an operator and the usability of
  %the system.

  %Methods: Participants replicated three construction models using Duplos.
  %Approximatively half of our participants (n=32) accomplished the task
  %alone and the other half accomplished the same task with a cobot (n=22).
  %We then used the NASA TLX to measure workload and, through a
  %sub-dimension, perceived cognitive load. Completion time (for each
  %model), number of errors (placing and replacing a piece), and the number
  %of gestures (movements of the upper limbs) were also measured.

  %Results: Collaboration with a cobot led to significantly fewer gestures
  %(51 vs. 74), fewer errors (2 vs. 8), and reduced perceived time demand,
  %but increased completion time (136 vs. 55 s). Perceived cognitive load
  %was not impacted by the cobot (36 vs. 37).

  %Conclusion: We conclude that collaboration with a cobot adapting to
  %human variability is possible, and that it could lead to better
  %performance and could improve certain dimensions of system usability.

\end{abstract}


\hypertarget{introduction}{%
\section{Introduction}\label{introduction}}

Collaborative robots (cobots) are increasingly popular in industrial
workplaces because they are perceived as a means to enhance performance,
particularly in future industrial workflows that routinely integrate new
technologies into production chains (INRS, 2018). The term ``cobot'',
created by Colgate \& collaborators (1996), refers to ``a robotic device
which manipulates objects in collaboration with a human operator'' (p.
1). The development of artificial intelligence offers cobots adaptive
qualities: adaptations to production dynamics, to technical
variabilities (Swamidass, 2000), and to the operator. Yet the artificial
algorithms remain at an experimental level and need to be improved and
evaluated before implanting cobots with adaptive capabilities into
industry. Human-cobot interactions can take on several forms. In this
report, we will only discuss direct collaboration (working
simultaneously on a task). To accept this collaboration, it is essential
to study the impacts of the introduction of collaborative robots on
operators before implementation (Kildal et al., 2018).

In many studies, the human-robot interaction experience is referred to
as ``user experience'' or ``UX'' (Hassenzahl, 2008). This experience can
be studied through the concept of usability (Bevan, 2009; Davis, 1989;
Dubey \& Rana, 2010; INRS, 2018; Nielsen, 1994). Indeed, studying a
system's usability is a way of guiding its design (Chaniaud et al.,
2020). According to ISO 9241-11, a standard reference defining
human-machine interaction (Bevan et al., 2015), usability is defined as
``the extent to which a product can be used by specified users to
accomplish specific goals with effectiveness, efficiency and
satisfaction in a particular context of use'' (ISO, 2018). Effectiveness
is the accuracy and degree of completion with which users will
accomplish set goals (ISO, 2018) and is measured through the quality of
the results, the quantity produced, and the percentage of goals
accomplished. In this study, we use the number of errors to measure
effectiveness. Indeed, this gives a behavioral measure of cognitive
load, since Verhulst (2018) defined cognitive load as the amount of
cognitive resources expended at a given time on a task.

Moreover, there are various measures for measuring cognitive load:
subjective and objective (Cain, 2007b; Kramer, 1990). Objective measures
can be obtained by studying individual performance on the task (Paas et
al., 2003): if a task is too onerous in cognitive resources, the
individual performing this task will no longer be able to provide all
the necessary resources and consequently their performance will
decrease. Efficiency is the degree to which the user performs the task
easily with a minimum of resources (ISO, 2018). We use task completion
time and the number of gestures to measure efficiency in our study.
Finally, satisfaction is defined as ``the extent to which the
user\textquotesingle s physical, cognitive and emotional responses
resulting from using a system meets the user\textquotesingle s needs and
expectations'' (ISO, 2018). The perceived cognitive load, an indicator
of the user's cognitive response as a result of using a system, is used
as a measure of the satisfaction dimension of usability. Subjective
measures of cognitive load can be obtained through questionnaires where
participants are asked questions directly. In this study, we use the
NASA-TLX (Hart \& Staveland, 1988), which measures workload as a whole
but also more specifically cognitive load, physical load, temporal
demands frustration, effort, and performance (Hart, 2006). Also, we
consider that usability and cognitive load are two theoretical fields
that could be linked to measure UX as a whole.

We hypothesized that a human-robot collaboration will reduce the
cognitive load required to perform a task, despite the potential
increase of cognitive load from introducing collaboration into the task
(Kildal et al., 2019). We assumed that collaboration will allow better
performance: increased effectiveness, efficiency, and satisfaction. Our
operational hypotheses were:

\begin{itemize}
\item
  H1 = perceived cognitive load will be lower for an assembly task in a
  condition of collaboration with the cobot than in a control condition.
\item
  H2 = completion time will be lower for an assembly task in a condition
  of collaboration with the cobot than in a control condition.
\item
  H3 = the number of errors will be lower for an assembly task in a
  condition of collaboration with the cobot than in a control condition.
\item
  H4 = the number of gestures will be lower for an assembly task in a
  condition of collaboration with the cobot than in a control condition.
\end{itemize}

\hypertarget{material-and-methods}{%
\section{Material and methods}\label{material-and-methods}}

54 participants (mean age = 21.4 years, SD = 3.9; 42 women and 12 men)
consented to participate in the experiment at the University of
Grenoble-Alpes. They did not receive any financial compensation for
their participation; however, psychology students could request extra
credits on a set of classes. They had no vision impairment
(colorblindness, etc.). Our study took place during the Covid
semi-lockdown and we obtained the approval of the director of the
inter-university laboratory of psychology of Grenoble to run the
experiment with human participants. Participants were given explanations
on the experiment itself and how their data will be handled. After this
introduction, they were given a consent form to sign in order to
continue the study. We developed a two-condition between-subject design.
Participants were randomly assigned to a group. In the control
condition, 32 participants performed the task alone, whereas in the
experimental condition, 22 participants performed the task in
collaboration with a cobot (Figure 1). The cobot used for the task is a
YuMi (short for You and Me) created by ABB Robotics enterprise (ASEA
Brown Boveri). It was used for three reasons: first, it is an industrial
cobot that is already implemented in many industrial factories, second,
it is the most ``collaborative'' robot currently existing, and third, it
has high safety included in the design (rubber padded joints, curved
angles, etc.) to work "safely" in close proximity with human operators.
Recruitment for the control condition took place in April because the
cobot was not ready yet. Recruitment for the experimental condition took
place in May, during the covid restriction. The two groups were uneven
for this reason.

\begin{figure}
\begin{center}
\includegraphics[width=2.18681in,height=2.10694in]{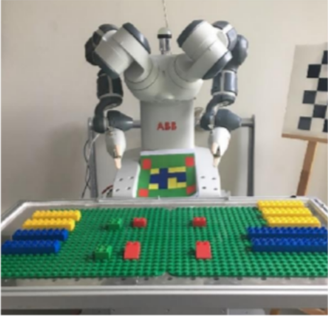}
\caption{The YuMi robot (ABB) used in the experimental condition.}
\end{center}
\end{figure}

The cobot was programmed on different levels through a modular robotic
architecture. First, the vision module perceives the environment and the
presence of the participant\textquotesingle s hand during the perception
phase to ensure a true description of the environment that is not
occluded by the participant\textquotesingle s hand. This stage was
performed using Python programming language, and the OpenCV computer
vision library. Second, the environment description obtained from the
perception module is used to infer the progress of the collaborative
task carried out by the robot and the participant. This information is
used to drive a cognitive conceptual model, which in turn makes use of
Automated Planning techniques to generate adaptive intelligent behavior
that corresponds to the variability of the assembly task. The
intelligent behavior is a plan: a sequence of actions that composes the
pick/place task (opening the gripper, moving the robot arm from one
place to another, closing the gripper, etc.). Third, the sequence of
actions generated to accommodate task variability in an intelligent
manner is executed on YuMi using the low-level assembly language RAPID
provided by ABB (the manufacturer of the robot). Finally, the plan
(intelligent behavior composed of a sequence of actions) and the robot
motion are projected on a screen through Graphical User Interface (GUI)
to explain the intelligent behavior to the participant by making them
aware of what will happen next.

Our aim was to develop a type of task that could be performed both in
collaboration with a cobot or alone, and that would support study of
cognitive load while considering the technical constraints of the cobot.
We chose to simulate a typical task in the industrial environment: an
assembly of components. In our exploratory study, the task was carried
out using Duplos (Lego). Each participant was required to replicate
three models (Figure 2).

\begin{figure}
\begin{center}
\includegraphics[width=3.12353in,height=1.28664in]{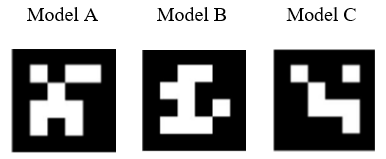}
\caption{The three models that participants were required to replicate.}
\end{center}
\end{figure}

The order of presentation of the models was randomized. Each participant
had a stock of Duplos located on their left (Area 1 in Figure 3). They
had to replicate the model in the ``construction zone'' (Area 2 in
Figure 3). In the experimental condition, the cobot\textquotesingle s
stock was placed on the participant\textquotesingle s right (Area 3 in
Figure 3).

\begin{figure}
\begin{center}
\includegraphics[width=3.08874in,height=1.74479in]{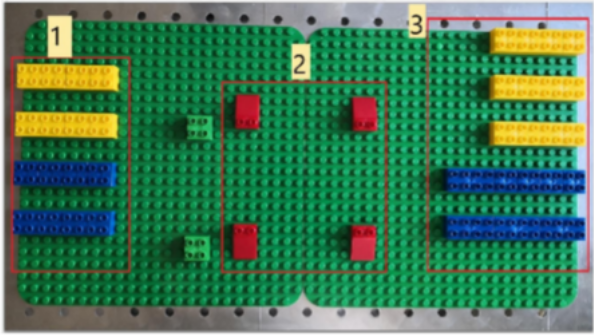}
\caption{Location of the construction zone (2), the participant's stock
(1), and the cobot's stock (3).}
\end{center}
\end{figure}

In the experimental condition, the cobot collaborated with the human by
placing pieces on the construction area at the same time as the
participant, by indicating if an error had been made using the
communication interface and by providing additional stock to the
participant (between the green Duplos left of Area 2). The cobot
alternated between actions (Table 1) and perception phases (during which
it would stop and analyze the Duplos placement). Its instructions were
the same as the human's: it had to complete the model. During most of
the experiment, the pace was set by the participant and the cobot had to
adapt. However, due to technical constraints, the participants could not
act when the cobot was in perception phase. They had to wait and regain
control of the pace once the perception phase ended.

\begin{table}
\caption{An example of a human-machine process chart in the experimental
condition.}

\begin{longtable}[]{@{}
  >{\raggedright\arraybackslash}p{(\columnwidth - 4\tabcolsep) * \real{0.2141}}
  >{\raggedright\arraybackslash}p{(\columnwidth - 4\tabcolsep) * \real{0.3937}}
  >{\raggedright\arraybackslash}p{(\columnwidth - 4\tabcolsep) * \real{0.3922}}@{}}
\toprule()
\begin{minipage}[b]{\linewidth}\raggedright
Time (in seconds)
\end{minipage} & \begin{minipage}[b]{\linewidth}\raggedright
Participant
\end{minipage} & \begin{minipage}[b]{\linewidth}\raggedright
Cobot
\end{minipage} \\
\midrule()
\endhead
0-5 & Looks at the model & Warns the participant of its next move on its
screen \\
5-10 & Looks at the screen to learn where the cobot will pick its piece
& Initiates the movement to pick a piece \\
10-15 & Picks a piece & Picks a piece \\
15-20 & Places a piece and looks at the model & Initiates the movement
to place the piece \\
20-25 & Picks and places two other pieces & Places a piece \\
25-30 & Picks and places another piece & Warns the participant of its
perception phase on its screen \\
30-40 & Waits and looks at the screen & Takes a picture of the
construction \\
40-45 & Looks at the model and the screen & Warns the participant of its
next move on its screen \\
45-50 & Picks and places another piece & Initiates the movement to pick
a piece to make up for a mistake made by the participant \\
50-55 & Picks and places two other pieces & Picks a piece \\
55-60 & Looks at the model & Initiates the movement to place the piece
in the correct space \\
60-65 & Picks and places the last piece & Places the piece in the
correct space \\
65-70 & Looks at the model and the construction & Warns the participant
of its perception phase on its screen \\
70-75 & Waits and looks at the screen & Takes a picture of the
construction \\
75-80 & Waits and looks at the screen and the model & Warns the
participant of the success of the construction \\
80-85 & Looks at the experimenter for instructions & Goes into standby
mode \\
\bottomrule()
\end{longtable}
\end{table}

In the control condition, participants had a three-minute training
session with a demonstration on how to complete a model. The experiment
took place in a ventilated room with adequate lighting. In the
experimental condition, participants had five minutes of training on how
to complete a model and how to interact with the cobot. No participant
demonstrated substantial mental or physical fatigue. They were given
special instructions to consult the screen of the cobot to know where it
was going to place its next piece and when it was going into the
perception phase.

All participants replicated the three models while being filmed. Video
recordings were analyzed using The Observer software (version 15.0.1200,
2019). We observed the completion time, number of errors (when a piece
is placed, removed, and replaced), and number of gestures, the latter
defined as movements of an upper limb (placing, removing a piece,
managing the stock, thinking, verification, waiting, robot interaction,
and prevented gestures).

Finally, participants completed a translated version (Cegarra \&
Morgado, 2009) of the NASA-TLX (Hart \& Staveland, 1988) to assess the
cognitive load they experienced (see Table 2). This test measures the
experienced mental load through six subscales: mental demand, physical
demand, time demand, frustration, effort. and performance. It is
conducted in two stages (Hart \& Staveland, 1988). First, we asked
participants to rate each subscale on a scale from 0 to 100. Second, we
asked them to rank the different subscales according to the impact of
each one on their overall feeling. This ranking ranges from 1 to 6, with
1 representing the subscale with the lowest impact on the overall
feeling and 6 representing the subscale with the highest impact on the
feeling of the mental load. All of this information allowed us to obtain
the weighted score of the cognitive load perceived by each participant
(each subscale is multiplied by the number of participants).

\begin{table}
\caption{French version of the NASA-TLX used in the experiment.}

\begin{longtable}[]{@{}
  >{\raggedright\arraybackslash}p{(\columnwidth - 2\tabcolsep) * \real{0.2496}}
  >{\raggedright\arraybackslash}p{(\columnwidth - 2\tabcolsep) * \real{0.7504}}@{}}
\toprule()
\begin{minipage}[b]{\linewidth}\raggedright
\emph{Mental demand}
\end{minipage} & \begin{minipage}[b]{\linewidth}\raggedright
Quelle quantité d'activité mentale et perceptive a été nécessaire pour
réaliser la tâche (par ex. réfléchir, décider, chercher, etc.) ? La
tâche vous a-t-elle paru simple, nécessitant peu d'attention, ou
complexe, nécessitant beaucoup d'attention ?
\end{minipage} \\
\midrule()
\endhead
\emph{Physical demand} & Quelle quantité d'activité physique a été
nécessaire pour réaliser la tâche (par ex. pousser, déplacer, tourner,
manipuler, etc.) ? La tâche vous a-t-elle paru facile, peu fatigante,
calme ou pénible, fatigante, active ? \\
\emph{Temporal demand} & Quelle pression temporelle avez-vous ressentie
durant l'exécution de la tâche ? Fallait-il gérer la réalisation de la
tâche de manière lente ou de manière rapide ? \\
\emph{Performance} & Comment estimez-vous votre performance en ce qui
concerne la réalisation de la tâche ? (Quel est votre niveau de
satisfaction concernant votre performance ?) \\
\emph{Effort} & Quel effort (mental et physique) avez-vous dû fournir
pour atteindre votre niveau de performance ? \\
\emph{Frustration} & Vous êtes-vous senti satisfait, content, relaxé ou
plutôt ennuyé, irrité, stressé pendant la réalisation de la tâche ? \\
\bottomrule()
\end{longtable}
\end{table}

The questionnaire was implemented using the QUALTRICS software
(Qualtrics, 2005). Internal consistency of the questionnaire was
confirmed: no item was correlated more than 0.70 with another, the KMO
index was greater than 0.50 (0.659), and the Bartlett Test was
significant (K=46.590(15), \emph{p}\textless0.01). Study results were
calculated using RStudio, version 1.4.1106 (RStudio Team, 2021). The
results presented below are from a dataset available on GitHub
(Datacobot2021a), and the specific open source code was applied
(RcodeDatacobot2021a). Unpaired \emph{t} tests were used to test the
study hypotheses, with \emph{p} \textless{} 0.05 considered
statistically significant. Assumptions of normal distributions of
equality of variances were tested, and the Wilcoxon test was used as
relevant.

\hypertarget{results}{%
\section{Results}\label{results}}

\hypertarget{hypothesis-1-the-effect-of-collaboration-with-a-cobot-on-perceived-cognitive-load}{%
\subsection{3.1 Hypothesis 1: the effect of collaboration with a cobot
on perceived cognitive
load}\label{hypothesis-1-the-effect-of-collaboration-with-a-cobot-on-perceived-cognitive-load}}

In the control condition, participants rated their cognitive load with
mean of 37.5 out of 100 (SD=12.9, min=15.2, max=59.1, N=32) and in the
experimental condition at 35.5 out of 100 (SD=15.3, min=11.3, max=62.5,
N=22). There was not significant effect of collaboration with the robot
on evaluation of cognitive load (\emph{t}=0.51\textsubscript{(52)},
\emph{p}=0.61).

\hypertarget{hypothesis-2-the-effect-of-collaboration-with-a-cobot-on-task-completion-time}{%
\subsection{3.2 Hypothesis 2: the effect of collaboration with a cobot
on task completion
time}\label{hypothesis-2-the-effect-of-collaboration-with-a-cobot-on-task-completion-time}}

The participants took an average of 54.8 seconds (SD=18.5, min=19.4,
max=146.4, N=32) to reproduce a model in the control condition and 136.4
seconds (SD=30.9, min=80.9 max=213.7, N=22) in the experimental
condition. There was a significant effect of collaboration with the
robot on the completion time (\emph{t}=-12.756\textsubscript{(52)},
\emph{p}=2,2*10\textsuperscript{-12}). Oddly, the NASA TLX answers
indicate that the participants perceived less temporal demand in the
collaboration condition (\emph{t=}4.6\textsubscript{(43.5)},
\emph{p=}3,668*10\textsuperscript{-5}).

\hypertarget{hypothesis-3-the-effect-of-collaboration-with-a-cobot-on-the-number-of-errors}{%
\subsection{3.3 Hypothesis 3: the effect of collaboration with a cobot
on the number of
errors}\label{hypothesis-3-the-effect-of-collaboration-with-a-cobot-on-the-number-of-errors}}

Participants made an average of eight errors (SD=5.8, min=0, max=20,
N=32) in the control condition and two errors (SD=1.8, min=0, max=6,
N=22) in the experimental condition. There was a significant effect of
collaboration with the robot on the number of errors (\emph{w}=618.5,
\emph{p}=2,556*10\textsuperscript{-6}).

\hypertarget{hypothesis-4-the-effect-of-collaboration-with-a-cobot-on-the-number-of-gestures-performed}{%
\subsection{3.4 Hypothesis 4: the effect of collaboration with a cobot
on the number of gestures
performed}\label{hypothesis-4-the-effect-of-collaboration-with-a-cobot-on-the-number-of-gestures-performed}}

The participants performed an average of 74 gestures (SD=20, min=41,
max=124, N=32) in the control condition and 51 (SD=9.8, min=34, max=72,
N=22) in the experimental condition. There was a significant effect of
collaboration with the robot on the number of gestures
(\emph{t}=4.9547\textsubscript{(52)},
\emph{p}=8,068*10\textsuperscript{-6}).

We tested the effect of the collaboration on all sub-sections of the
NASA TLX test. T-tests revealed no differences on those sub-sections
(including perceived frustration and effort), with the exception of a
positive significant effect of the collaboration on the perceived time
demand.

\hypertarget{discussion}{%
\section{Discussion}\label{discussion}}

The current results concerning the reduction of the number of errors and
the number of gestures in the collaboration condition are in line with
our initial hypothesis (H3 and H4). In our study, the human/robot
collaboration reduced the number of errors and the number of gestures
performed. This outcome could imply that in an non-experimental
industrial work task, collaboration with a cobot could improve operator
efficiency and the quality of their work. Similar results have been
reported earlier. Indeed, Salunkhe et al. (2019) recently found that
cobots could, in theory, identify problematic gestures, thus protecting
human workers, while improving the quality of the work.

Since introducing the cobot increased the number of instructions
(participants had to understand the task, and understand how the cobot
works and how to interact with it), we feared that those additional
instructions would adversely impact perceived cognitive load. However,
this was not the case, since the perceived mental loads did not differ
in the two conditions and H1 was not supported, though collaboration
with the cobot had no negative effect on perceived cognitive load.
Kildal et al. (2019) found that the complexity of the task is not
increased due to the introduction of a collaboration between human and
cobot, even when the operator is cognitively impaired. Future studies
need to measure the impact of collaborating with a cobot on cognitive
load, as it is representative of the operator's wellbeing in their work
(Fruggiero et al., 2020).

Finally, the results concerning completion time (H2) were against our
expectations. Indeed, we found that completion time significantly
increased in the condition of co-activity. This increase was due to the
introduction of waiting time and technical problems that occurred during
the experiment. In fact, another study, realized in an industrial
context, showed no increase in completion time between collaboration
with a cobot and no collaboration (Fager et al., 2019). Nonetheless, the
perceived time demand was significantly lower here in the experimental
condition. Thus, even though collaboration with the cobot lasted longer,
this was not reflected in the perceptions of the participants. This
outcome could be a clue that perceived pleasure with a cobot is higher
that the perceived pleasure of working alone (El Makrini et al., 2018).
Overall, we concluded that collaboration improved the user experience on
the following criteria: errors, gestures and perceived time demand
(Table 3).

\begin{table}
\caption{Summary of the results of collaborating with a cobot for each
independent variable.}

\begin{longtable}[]{@{}
  >{\raggedright\arraybackslash}p{(\columnwidth - 2\tabcolsep) * \real{0.4313}}
  >{\raggedright\arraybackslash}p{(\columnwidth - 2\tabcolsep) * \real{0.5687}}@{}}
\toprule()
\begin{minipage}[b]{\linewidth}\raggedright
\textbf{Outcome}
\end{minipage} & \begin{minipage}[b]{\linewidth}\raggedright
\textbf{Collaboration with cobot}
\end{minipage} \\
\midrule()
\endhead
\begin{minipage}[t]{\linewidth}\raggedright
Perceived cognitive load
\end{minipage} & \begin{minipage}[t]{\linewidth}\raggedright
\textbf{Equal} to the control condition
\end{minipage} \\
\begin{minipage}[t]{\linewidth}\raggedright
Completion time
\end{minipage} & \begin{minipage}[t]{\linewidth}\raggedright
\textbf{Higher} than in the control condition
\end{minipage} \\
\begin{minipage}[t]{\linewidth}\raggedright
Number of errors
\end{minipage} & \begin{minipage}[t]{\linewidth}\raggedright
\textbf{Lower} than in the control condition
\end{minipage} \\
\begin{minipage}[t]{\linewidth}\raggedright
Number of gestures
\end{minipage} & \begin{minipage}[t]{\linewidth}\raggedright
\textbf{Lower} than in the control condition
\end{minipage} \\
\begin{minipage}[t]{\linewidth}\raggedright
Perceived time demand
\end{minipage} & \begin{minipage}[t]{\linewidth}\raggedright
\textbf{Lower} than in the control condition
\end{minipage} \\
\bottomrule()
\end{longtable}
\end{table}

This study provides several contributions regarding human-machine
systems design. Indeed, we noticed that there are variabilities in human
behavior during the realization of a laboratory-simulated assembly task.
These variabilities emerged from organizational design (presence or
absence of a cobot), technical variability (issues linked to the cobot
on certain experimental runs), and inter-individual variabilities (e.g.
wide range of the number of gestures performed by participants).

These variabilities were managed within the collaboration, as was the
case in other recent studies (Chacón et al., 2020; Fager et al., 2019).
The cobot and its AI allowed the experimental task to be constrained in
terms of how the assembly was done (what piece to put, the order of
piece, etc.) only by the human participant and not by ``the machine''.
Our work shows that it is possible to have an intelligent machine that
regulates its actions to help the human, whatever strategy he adopts,
which is an advantage for industry work (Paulikov{\'a} et al., 2021).
Furthermore, the current work showed that this new type of collaboration
could be beneficial in terms of performance and usability. Further
studies in an industrial setting, though, are needed to confirm such
benefits. Those studies need to address the impacts (negative and
positive) that cobots have on the operators (well-being, health,
satisfaction at work) and on the industry (performance, security), the
protocol of implementation used to change the organization of work, how
operators are trained and recruited to work with cobots, and the ethics
behind those changes. There needs to be an emphasis on the study of
social aspects of the human/cobot collaboration.

We have also verified that it is possible (in the context of a
laboratory task) to measure cognitive load through different behavioral
measures and to use these same indicators to obtain information on
usability criteria according to the ISO 9241-11. Investigating cognitive
load needed to adapt cobots to the human is a new, challenging, and
necessary field of research (Fruggiero et al., 2020). It is, therefore,
an original contribution to bring together these two concepts belonging
to different disciplinary fields and by experimenting in a different
domain of UX. We offer an original experimental protocol that can be
transposed to other types of tasks.

The cobot had technical problems that impacted some experimental runs.
By working on improving of the cobot's technical capacities, we will be
able to reduce task completion time and thus improve the overall UX.
Furthermore, it would be interesting to replicate this study in an
industrial environment, which is more complex than laboratory-controlled
situations. The task here was quite simple, thus limiting the
generalizability of the study. However, this simplicity was necessary to
improve the AI of the cobot, a first step to ensure that AI was able to
adapt to the variabilities of the human.

Our work indicates that the adaptation of a robotic system to a human is
possible and should receive priority consideration owing to its
benefits: it requires very little training (since the human does not
adapt to the cobot) and it is flexible to human variabilities. Also, no
analysis was made regarding potential differences related to gender, as
there were more women than men in our study (statistical analysis would
have been biased). Future studies should address gender (Perez, 2019) as
a potential source of interindividual variability as it is relevant to
improve science and design (Tannenbaum et al., 2019).

An experiment with a within-subjects design, by randomizing the order of
the two conditions, would allow us to assess the perception of
difference between the two conditions. Indeed, participant behaviors
were different across experimental conditions. For example, the
participants were much more static in the experimental condition than in
the control condition. This difference in behavior could be the result
of a different conceptualization of the task.

Future research needs to be developed to evaluate the impact of the
cobot collaboration on workload. For now, our study is in line with
earlier literature, and it seems that using a cobot to collaborate with
an operator could decrease perceived cognitive load. As Perre et al.
(2018) suggested in their article, the cobot could ``diminish an
operator's workload and improve the working conditions'' (p. 5). To
analyze this impact, researchers can use self-reported scales such as
the NASA-TLX and behavioral measurements such as the number of errors
and the completion time.

In summary, we found that performing an assembly task in collaboration
with a cobot increased effectiveness, maintained the level of
satisfaction, and improved certain criteria of efficiency (number of
errors, number of gestures) if the robot is programmed to adapt its
decisions to the human's. In this laboratory context, the human/cobot
collaboration facilitated better task performance and an overall
improved UX. We also developed an experimental protocol linking the
concepts of usability and cognitive load in a multidisciplinary
experimental task, combining the technical constraints of programmers
and the constraints of human behavior measurements. This protocol can be
used in future studies to identify key points to consider when designing
a cobot (perception of the cobot, infrastructure of the AI, etc.), thus
facilitating broader and more effective implementation of cobots in
Industry 4.0.

\hypertarget{references}{%
\section{References}\label{references}}

Bevan, N. (2009). What is the difference between the purpose of
usability and user experience evaluation methods. \emph{Proceedings of
the Workshop UXEM}, \emph{9}(1), 1--4.

Bevan, N., Carter, J., \& Harker, S. (2015). ISO 9241-11 Revised: What
Have We Learnt About Usability Since 1998? In M. Kurosu (Ed.),
\emph{Human-Computer Interaction: Design and Evaluation} (pp. 143--151).
Springer International Publishing.
https://doi.org/10.1007/978-3-319-20901-2\_13

Cain, B. (2007a). \emph{A Review of the Mental Workload Literature}.
Defense Research and Development Toronto.
https://www.semanticscholar.org/paper/A-Review-of-the-Mental-Workload-Literature-Cain/a41e553eea5621e8b20b4ad6fe9ab5bcab02a284

Cegarra, J., \& Morgado, N. (2009). \emph{Étude des propriétés de la
version francophone du NASA-TLX}. COOP'2000.
https://docplayer.fr/21755956-Etude-des-proprietes-de-la-version-francophone-du-nasa-tlx.html

Chacón, A., Ponsa, P., \& Angulo, C. (2020). On Cognitive Assistant
Robots for Reducing Variability in Industrial Human-Robot Activities.
\emph{Applied Sciences}, \emph{10}(15), 5137.
https://doi.org/10.3390/app10155137

Chaniaud, N., Métayer, N., Megalakaki, O., \& Loup-Escande, E. (2020).
Effect of Prior Health Knowledge on the Usability of Two Home Medical
Devices: Usability Study. \emph{JMIR MHealth and UHealth}, \emph{8}(9),
e17983. https://doi.org/10.2196/17983

Colgate, J. E., Wannasuphoprasit, W., \& Peshkin, M. A. (1996).
\emph{Cobots: Proceedings of the 1996 ASME International Mechanical
Engineering Congress and Exposition}. 433--439.
http://www.scopus.com/inward/record.url?scp=0030402971\&partnerID=8YFLogxK

Davis, F. D. (1989). Perceived Usefulness, Perceived Ease of Use, and
User Acceptance of Information Technology. \emph{MIS Quarterly},
\emph{13}(3), 319--340. https://doi.org/10.2307/249008

Dubey, S., \& Rana, A. (2010). Analytical roadmap to usability
definitions and decompositions. \emph{International Journal of
Engineering Science and Technology}, \emph{2}(9), 4723--4729.

El Makrini, I., Elprama, S. A., Van den Bergh, J., Vanderborght, B.,
Knevels, A.-J., Jewell, C. I. C., Stals, F., De Coppel, G., Ravyse, I.,
Potargent, J., Berte, J., Diericx, B., Waegeman, T., \& Jacobs, A.
(2018). Working with Walt: How a Cobot Was Developed and Inserted on an
Auto Assembly Line. \emph{IEEE Robotics \& Automation Magazine},
\emph{25}(2), 51--58. https://doi.org/10.1109/MRA.2018.2815947

Fager, P., Calzavara, M., \& Sgarbossa, F. (2019). Kit Preparation with
Cobot-supported Sorting in Mixed Model Assembly.
\emph{IFAC-PapersOnLine}, \emph{52}(13), 1878--1883.
https://doi.org/10.1016/j.ifacol.2019.11.476

Fruggiero, F., Lambiase, A., Panagou, S., \& Sabattini, L. (2020).
Cognitive Human Modeling in Collaborative Robotics. \emph{Procedia
Manufacturing}, \emph{51}, 584--591.
https://doi.org/10.1016/j.promfg.2020.10.082

Hart, S. G. (2006). \emph{NASA-task load index (NASA-TLX); 20 years
later}. \emph{50}, 904--908.

Hart, S. G., \& Staveland, L. E. (1988). Development of NASA-TLX (Task
Load Index): Results of Empirical and Theoretical Research. In P. A.
Hancock \& N. Meshkati (Eds.), \emph{Advances in Psychology} (Vol. 52,
pp. 139--183). North-Holland.
https://doi.org/10.1016/S0166-4115(08)62386-9

Hassenzahl, M. (2008). User experience (UX): Towards an experiential
perspective on product quality. \emph{Proceedings of the 20th Conference
on l'Interaction Homme-Machine}, 11--15.
https://doi.org/10.1145/1512714.1512717

INRS. (2018). \emph{Dossier Robots collaboratifs}.
www.inrs.fr/risques/robots-collaboratifs.html

Kildal, J., Martín, M., Ipiña, I., \& Maurtua, I. (2019). Empowering
assembly workers with cognitive disabilities by working with
collaborative robots: A study to capture design requirements.
\emph{Procedia CIRP}, \emph{81}, 797--802.
https://doi.org/10.1016/j.procir.2019.03.202

Kildal, J., Tellaeche, A., Fernández, I., \& Maurtua, I. (2018).
Potential users' key concerns and expectations for the adoption of
cobots. \emph{Procedia CIRP}, \emph{72}, 21--26.
https://doi.org/10.1016/j.procir.2018.03.104

Kramer, A. (1990). \emph{Physiological Metrics of Mental Workload: A
Review of Recent Progress}. https://doi.org/10.21236/ada223701

Nielsen, J. (1994). \emph{Usability Engineering}. Morgan Kaufmann
Publishers Inc.

Paas, F., Tuovinen, J. E., Tabbers, H., \& Van Gerven, P. W. M. (2003).
Cognitive Load Measurement as a Means to Advance Cognitive Load Theory.
\emph{Educational Psychologist}, \emph{38}(1), 63--71.
https://doi.org/10.1207/S15326985EP3801\_8

Paulikov{\'a},A.;Gyur{\'a}k Babel'ov{\'a}, Z.; Ub{\'a}rov{\'a}, M. (2021) Analysis of the
Impact of Human--Cobot Collaborative Manufacturing Implementation on the
Occupational Health and Safety and the Quality Requirements. \emph{Int.
J. Environ. Res. Public Health} 2021, \emph{18}, 1927. https://
doi.org/10.3390/ijerph18041927

Perez, C. C. (2019). Invisible women: Data bias in a world designed for
men. Abrams.

Perre, G. V. de, Makrini, I. E., Acker, B. B. V., Saldien, J., Vergara,
C., Pintelon, L., Chemweno, P., Weuts, R., Moons, K., Dewil, R.,
Cattrysse, D., Burggraeve, S., Mateus, J. C., Decré, W., Aertbeliën, E.,
\& Vanderborght, B. (2018, October). \emph{Improving productivity and
worker conditions in assembly: Part 1 - A collaborative architecture and
task allocation}. 2018 IEEE/RSJ International Conference on Intelligent
Robots and Systems, IROS 2018: Towards a Robotic Society.
https://research.utwente.nl/en/publications/improving-productivity-and-worker-conditions-in-assembly-part-1-a

Tannenbaum, C., Ellis, R. P., Eyssel, F., Zou, J., \& Schiebinger, L.
(2019). Sex and gender analysis improves science and engineering.
Nature, 575(7781), 137-146.
doi:http://dx.doi.org/10.1038/s41586-019-1657-6

Salunkhe, O., Stensöta, O., Åkerman, M., Berglund, Å. F., \& Alveflo,
P.-A. (2019). Assembly 4.0: Wheel Hub Nut Assembly Using a Cobot.
\emph{IFAC-PapersOnLine}, \emph{52}(13), 1632--1637.
https://doi.org/10.1016/j.ifacol.2019.11.434

Swamidass, P. M. (Ed.). (2000). Manufacturing in variability. In
\emph{Encyclopedia of Production and Manufacturing Management} (pp.
825--825). Springer US. https://doi.org/10.1007/1-4020-0612-8\_1032

Verhulst, E. (2018). \emph{Contribution de l'étude de l'interaction en
environnement virtuel: Intérêt de la charge mentale} {[}Phdthesis,
Université d'Angers{]}. https://tel.archives-ouvertes.fr/tel-02157566

\end{document}